\documentstyle[12pt,epsfig,cite]{elsart}

\newcommand{\fins}{D^{(*)} K^- K^{(*)0}}
\newcommand{\allc}{B \rightarrow D^{(*)} K^- K^{(*)0}}
\newcommand{\doko}{B^- \rightarrow D^0 K^- K^{0}_{S}}
\newcommand{\dpko}{\bar{B^0} \rightarrow D^+ K^- K^{0}_{S}}
\newcommand{\dsoko}{B^- \rightarrow D^{*0} K^- K^{0}_{S}}
\newcommand{\dspko}{\bar{B^0} \rightarrow D^{*+} K^- K^{0}_{S}}
\newcommand{\dokoo}{B^- \rightarrow D^0 K^- K^{0}}
\newcommand{\dpkoo}{\bar{B^0} \rightarrow D^+ K^- K^{0}}
\newcommand{\dsokoo}{B^- \rightarrow D^{*0} K^- K^{0}}
\newcommand{\dspkoo}{\bar{B^0} \rightarrow D^{*+} K^- K^{0}}
\newcommand{\dokst}{B^- \rightarrow D^0 K^- K^{*0}}
\newcommand{\dpkst}{\bar{B^0} \rightarrow D^+ K^- K^{*0}}
\newcommand{\dsokst}{B^- \rightarrow D^{*0} K^- K^{*0}}
\newcommand{\dspkst}{\bar{B^0} \rightarrow D^{*+} K^- K^{*0}}
\newcommand{\kmko}{K^-K^{0}_{S}}
\newcommand{\kmkst}{K^-K^{*0}}

\begin{document}

\begin{frontmatter}

 \begin{flushleft}
 \vspace*{1.cm}
 \epsfysize3cm
 \epsfbox{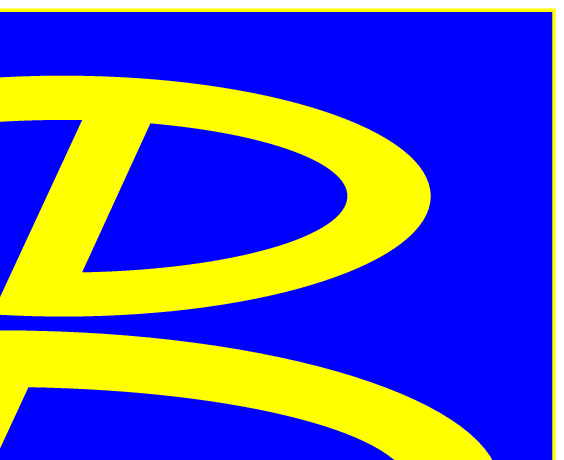}    
 \vspace*{1.cm}
 \end{flushleft}

 \begin{flushright}
 \large
 \vspace*{-4.cm}
 \hspace*{4.in}\mbox{Belle Preprint 2002-25} \\
 \vspace*{0.3cm}
 \hspace*{4.in}\mbox{KEK Preprint 2002-72} \\
 \vspace*{2.cm}
 \end{flushright}



\title{
\Large \bf Observation of {\boldmath{$B$}} $\rightarrow$ {\boldmath{$D$}}$^{(*)}${\boldmath{$K$}}$^-${\boldmath{$K$}}$^{0(*)}$ decays
}


  \collab{Belle Collaboration}
\begin{center}
  \author[ITEP]{A.~Drutskoy}, 
  \author[KEK]{K.~Abe}, 
  \author[TohokuGakuin]{K.~Abe}, 
  \author[TIT]{N.~Abe}, 
  \author[Niigata]{R.~Abe}, 
  \author[Tohoku]{T.~Abe}, 
  \author[KEK]{I.~Adachi}, 
  \author[Korea]{Byoung~Sup~Ahn}, 
  \author[Tokyo]{H.~Aihara}, 
  \author[Nagoya]{M.~Akatsu}, 
% ...
\end{center}

% ...
\address[KEK]{High Energy Accelerator Research Organization (KEK), Tsukuba, Japan}
% ...
\address[ITEP]{Institute for Theoretical and Experimental Physics, Moscow, Russia}
% ...
\address[Korea]{Korea University, Seoul, South Korea}
% ...
\address[Nagoya]{Nagoya University, Nagoya, Japan}
% ...
\address[Niigata]{Niigata University, Niigata, Japan}
% ...
\address[TohokuGakuin]{Tohoku Gakuin University, Tagajo, Japan}
\address[Tohoku]{Tohoku University, Sendai, Japan}
\address[Tokyo]{University of Tokyo, Tokyo, Japan}
\address[TIT]{Tokyo Institute of Technology, Tokyo, Japan}
% ...
\thanks[NovaGorica]{on leave from Nova Gorica Polytechnic, Nova Gorica, Slovenia}

\maketitle

\normalsize

\begin{abstract}

The $\allc$ decays have been observed for the first time.
The branching fractions of the $\allc$ decay modes are measured.
Significant signals are found for the $B \rightarrow D^{(*)} K^- K^{*0}$
and $B^- \rightarrow D^0 K^- K^{0}_{S}$ decay modes.
The invariant mass and polarization distributions for the $\kmkst$
and $\kmko$ subsystems have been studied.
For the $\kmkst$ sybsystem these distributions agree well with
those expected for two-body $B \rightarrow D^{(*)} a_1^-(1260)$ decays,
with $a_1^-(1260) \rightarrow K^- K^{*0}$.
The analysis was done using 29.4 fb$^{-1}$ of data collected 
with the Belle detector at the $e^+ e^-$ asymmetric collider KEKB.
\end{abstract}



\setcounter{footnote}{0}

\end{frontmatter}

\newpage

\normalsize

\section{Introduction}

In this paper we report on the first observation of 
$B$ meson decays to \linebreak $\fins$ final 
states\footnote{Charge conjugate modes are implicitly included.}.
Such decays require the creation of an additional $s\bar{s}$ pair and 
can occur via the quasi-two-body mechanism shown in Fig.~1a, where 
the intermediate resonance 
decays to $K^- K^{0(*)}$, or by a non-resonant three-body decay shown in 
Fig.~1b.
Contributions from color-suppressed diagrams are expected to be 
small \cite{bsw}.
The investigation of these processes can provide important 
information for testing various models of hadronic $B$ decays
as well as for studies of $K^- K^{0(*)}$ resonant states.

\begin{figure}[h!]
\vspace{-0.9cm}
\begin{center}
\hspace{-0.8cm}
\epsfig{file=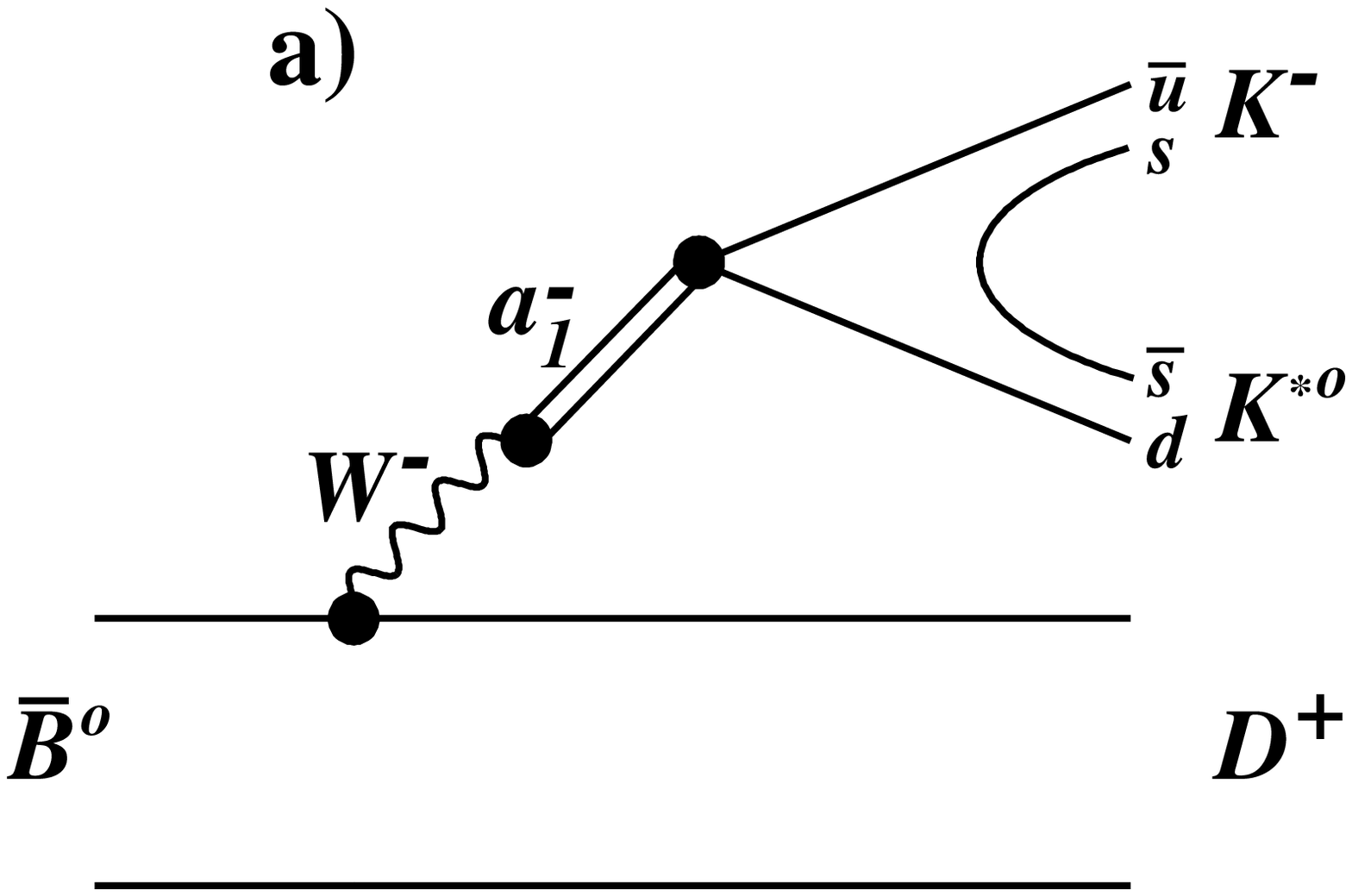,width=7cm,height=7cm}\epsfig{file=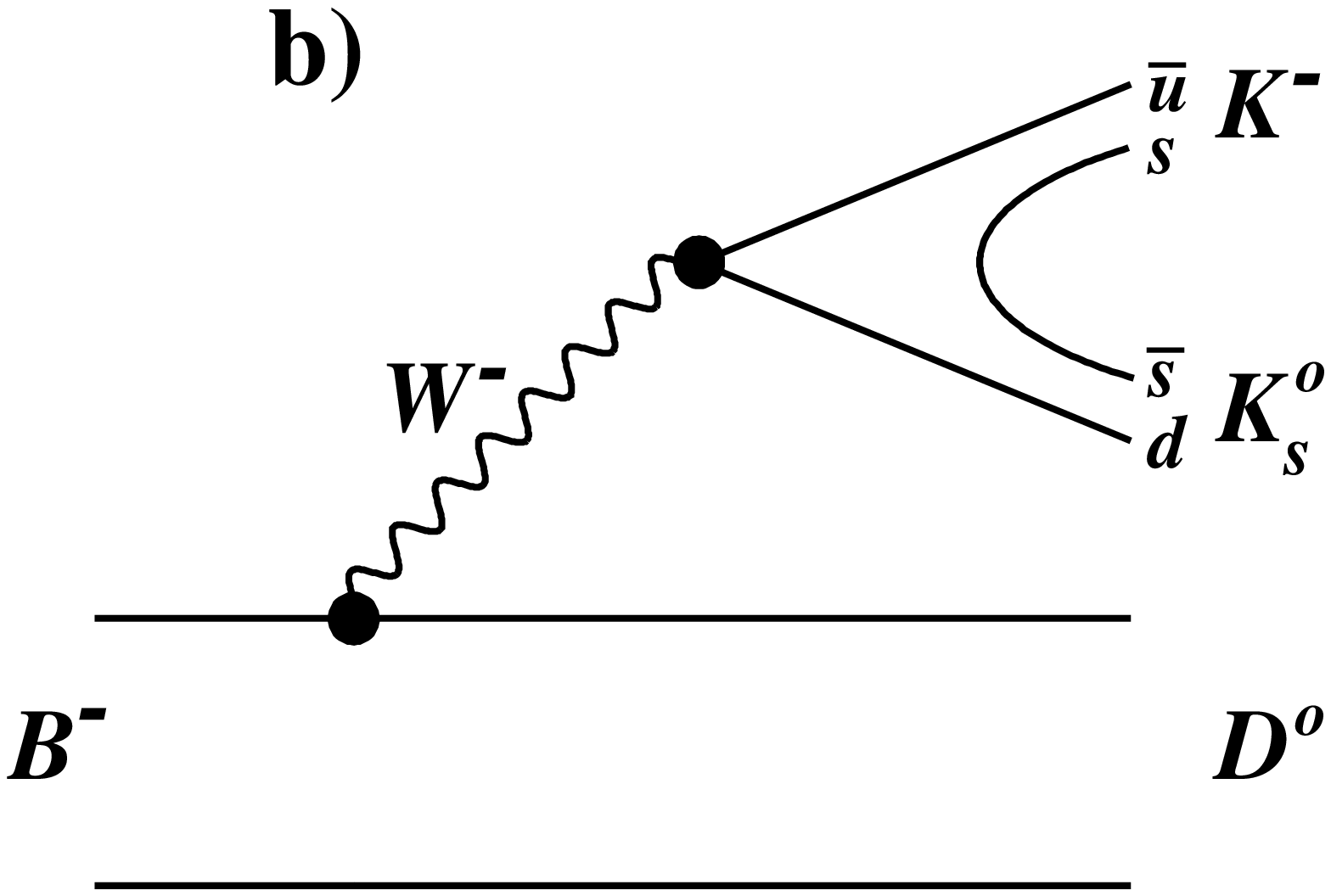,width=7cm,height=7cm}
\end{center}
\vspace{-1.4cm}
\caption{ The external spectator diagrams for a) quasi-two-body and 
b) non-resonant three-body $B$ meson decays with the $s\bar{s}$ pair 
creation. The decay channels $\dpkst$ and $\doko$ are used as examples.}
\label{diagramms}
\end{figure}

In $\allc$ decays, any
intermediate $K^- K^{0(*)}$ resonance must have isospin 1. The 
allowed quantum numbers are J$^P$=$\,0^-$,$1^-$,$1^+$ for 
the $K^- K^{*0}$ final state and 
J$^P$=$\,1^-$ for the $K^- K^0_S$ final state.
A $K^- K^0_S$ state with J$^P$=$\,0^+$
cannot be produced via tree diagrams of Figs.~1a,b \cite{aofor}
in the limit of exact isospin symmetry.
The production of 
resonances with spin larger than one is forbidden in the factorization 
approach \cite{bsw}. Furthermore, these restrictions on the quantum numbers
for $K^- K^{0(*)}$ resonant states also hold for
non-resonant $K^- K^{0(*)}$ production (Fig.~1b) \cite{aofor}.

In the case of decays via intermediate resonances,
the final state branching fraction depends on both
the production and decay rates of the resonance.
Large production rates of the order of $1\%$ have been observed 
for the $B \rightarrow D^{(*)} a_1(1260)$
and $B \rightarrow D^{(*)} \rho(770)$ decay channels \cite{cleoa}.
The $\kmkst$ final state can be produced in the decay of the
$a_1^-(1260)$ resonance,
however, the branching fraction for $a_1^-(1260) \rightarrow \kmkst$,
which was indirectly determined in $\tau$ decays by the CLEO 
collaboration \cite{cleob}, was found to be 
only $(3.3\,\pm\,0.5)\%$. One also expects that for
the J$^P$=$1^-$ resonances $\rho(1450)$ and $\rho(1700)$ 
the branching fractions to $K^-K^{0(*)}$ are at the level of a few percent
or less \cite{pdg}. For $\kmko$ production in $\rho(770)$ 
meson decay the phase space is very limited.

In contrast to two-body decays, non-resonant three-body $B$ decays 
to one charmed and two light mesons have not yet been observed \cite{cleoa}.
It should be noted that $s\bar{s}$ pair creation in a
non-resonant three-body $B$ decay (Fig.~1b) is only moderately suppressed 
relative to that for $u\bar{u}$ and $d\bar{d}$
pairs if SU(3) and phase space effects are taken into account \cite{Alek}.
Unfortunately, quasi-two-body and three-body processes 
cannot be easily separated because of the large widths of
the intermediate resonances.

The branching fractions for $D$ decays with $s\bar{s}$ creation 
are measured to be \mbox{$\sim(1-2)\%$} \cite{pdg}.
However, these values cannot be directly extrapolated to
$B$ decay because final state interactions,
which are much larger for $D$ decays,
can significantly affect the decay rates.
A possible similarity between the diagrams of Fig.~1 and those 
describing $\tau$ lepton decays with $K^-K^{0(*)}$ in the final state is
obscured by the rather limited phase space, resulting in  
small branching fractions for $\tau$ decays to
$K^- K^0 \nu_{\tau}$ and $K^- K^{*0} \nu_{\tau}$ when compared to 
non-$s\bar{s}$ channels.

In this paper the branching fractions for $\allc$ decays
are obtained for the full allowed kinematic region.
The invariant mass and polarization distributions of the $K^-K^{0(*)}$ 
subsystem are then studied in detail.

\section{Belle Detector}

The data were collected with the Belle detector at KEKB,
an asymmetric energy double storage ring collider
with 8\,GeV electrons and 3.5\,GeV positrons \cite{kekb}.
The results are based upon a data sample with an integrated luminosity of 
29.4\,fb$^{-1}$ taken at the $\Upsilon$(4S) resonance which
corresponds to 31.9 million $B\bar{B}$ pairs.

Belle is a general-purpose detector with a 1.5 T superconducting
solenoid magnet. Charged particle tracking 
is provided by a silicon vertex detector (SVD), consisting
of three nearly cylindrical layers of double-sided
silicon strip detectors, and a 50-layer central drift chamber (CDC).
Particle identification is accomplished by combining the information
from silica aerogel \v{C}erenkov counters (ACC) and a 
time-of-flight counter system (TOF) with 
specific ionization ($dE/dx$) measurements in the CDC.
The combined response of the three
systems provides at least 2.5$\sigma$ equivalent $K/\pi$ separation
for laboratory momenta up to 3.5\,GeV/c.
A CsI(Tl) electromagnetic calorimeter (ECL) located inside
the solenoid coil is used for detection of photons and electrons.
The detector is described in detail elsewhere \cite{BELLE_DETECTOR}.

The Monte Carlo (MC) simulation event samples used in this analysis 
were generated using a detailed GEANT-based simulation of the Belle detector 
response. 
Simulated events were processed in a manner similar to the data.

\section{Selection Criteria}

The charged track momenta are reconstructed using the 
CDC and SVD tracking system.
Kaon and pion mass hypotheses are assigned to charged 
tracks 
using a likelihood ratio $P(K/\pi) = L(K) / (L(K)+L(\pi))$, which ranges
between 0 and 1. $P(K/\pi)>\,0.6$ is required for kaon 
candidates and $P(K/\pi)<\,0.6$ for pion candidates.

Candidate $\pi^0$ mesons are reconstructed from pairs of photons
in the ECL with invariant masses 
within $\pm 15\,$MeV/c$^2$ of the nominal $\pi^0$ mass
($\sim$ 3$\sigma$ in the $\pi^0$ mass resolution).
The $\pi^0$ daughter photons are required to have energies greater
than 30\,MeV. A mass-constrained kinematic fit is performed
on the $\pi^0$ candidates to improve their energy resolution.

$K^0_S$ candidates are formed from $\pi^+\pi^-$ combinations
with an invariant mass within $\pm 10\,$MeV/c$^2$ of the nominal
$K^0_S$ mass ($\sim$ 3$\sigma$).
The two pions are required to have a common vertex 
that is displaced from the interaction point by at least 0.4\,cm in the
plane perpendicular to the beam direction. The difference
in $z$ coordinates (parallel to the beam direction) for the tracks
constituting the secondary vertex is required to be less than 2\,cm.
The angle $\alpha$ between 
the $K^0_S$ flight direction and the measured $K^0_S$ momentum direction
is required to satisfy cos$\,\alpha\,>0.8$.

An opposite sign $K$ and $\pi$ meson combination is taken 
as a $K^{*0}$ candidate if its invariant mass lies within a 
$\pm 50\,$MeV/c$^2$ interval of the nominal $K^{*0}$ mass.

The decay channels $D^+\rightarrow K^-\pi^+\pi^+$ and 
$D^0\rightarrow K^-\pi^+$ are used in this analysis for all 
studied $B$ decay modes, in order to avoid large combinatorial background.
The decay channel $D^0\rightarrow K^-\pi^-\pi^+\pi^+$ is only added
for the $D^{*0}$ and $D^{*+}$ meson reconstruction, because the $D^*$
constraint strongly suppresses the combinatorial background.
The invariant mass of the $D$ candidates is required 
to lie within $\pm 15\,$MeV/c$^2$ of the nominal $D$ mass 
for the first two modes
and within $\pm 12\,$MeV/c$^2$ for the $K^-\pi^-\pi^+\pi^+$ mode
($\sim$ 3$\sigma$).
A mass and vertex constrained kinematic fit is then performed
on the $D$ candidates and results of good quality fits 
are used to improve the $D$ momentum resolution.

For $D^{*0}$ and $D^{*+}$ candidates, the
$D^{*0} \rightarrow D^0\pi^0$ and $D^{*+} \rightarrow D^0\pi^+$
decay modes are used. The invariant mass of the
$D^0$ and $\pi$ combination is required to be
within $\pm 3.0\,$MeV/c$^2$ of
the nominal value for $D^{*0}$ ($\sim$ 3$\sigma$)  
and within \mbox{$\pm 1.5$ MeV/c$^2$} for $D^{*+}$ ($\sim$ 3$\sigma$).

The $D$ and $D^*$ candidates are combined with the $K^- K^0_S$ and
$K^- K^{*0}$ candidates to form $B^-$ and $\bar{B}^0$ candidates.
Two kinematic variables are used to extract the $B$ meson signal,
the energy difference $\Delta E\,=\,E^{\rm CM}_B-E^{\rm CM}_{\rm beam}$ and
the beam-constrained mass $M_{\rm bc}=\sqrt{(E^{\rm CM}_{\rm beam})^2\,-\,(p^{\rm CM}_B)^2}$,
where $E^{\rm CM}_B$ and $p^{\rm CM}_B$ are the center of mass (CM) energy and momentum
of the $B$ candidate and $E^{\rm CM}_{\rm beam}$ is the CM beam energy.
The intervals $M_{\rm bc} > 5.2$\,GeV/c$^2$ and $|\Delta E|\,<0.2$\,GeV
are selected.
The MC estimate of the $M_{\rm bc}$ resolution
is in the range (2.7\,--\,3.0)\,MeV/c$^2$;
the $\Delta E$ resolution ranges between 12.5 and 14.5\,MeV for final
states including $\pi^0$'s and between 8.5 and 10.5\,MeV for 
other final states.
The background from $B \rightarrow D^{(*)} D_s^-$ decay modes,
with $D_s^- \rightarrow K^-K^{0(*)}$, 
is removed by requiring 
\mbox{$|M(D_s^-) - M(K^-K^{0(*)})| > 20$\,MeV/c$^2$ ($\sim$ 4$\sigma$)}.
Only one $B$ meson candidate per event is accepted. 
In cases of multiple entries, the chosen candidate is the one with
the largest likelihood value, which is calculated using information
about the differences between nominal and measured
$D$, $D^*$ and $\pi^0$ masses and the charged particle identification.

The continuum background under the $\Upsilon$(4S) signal is suppressed
by topological cuts that were optimized using
MC to model the signal and data in the $B$ mass sideband 
$(5.2\,<\,M_{\rm bc}\,<\,5.26)\,$GeV/c$^2$ to model background. 
The ratio of the second to the zeroth Fox-Wolfram moments \cite{fox}
is required to be less than 0.5. The angle $\theta_{thr}^*$ in the CM
between the thrust axes of the particles forming the $B$ candidate and 
all other particles in the event must satisfy
$|\rm{cos}\it{(\theta_{thr}^*)}|<$ 0.85.
The angle $\theta_{B}^*$ in the CM between the beam direction and the
$B$ momentum direction should lie in the range
$|\rm{cos}\it{(\theta_{B}^*)}|<$ 0.9.

\section{Systematics and Backgrounds} 

 The systematic uncertainties described below were estimated and
added in quadrature to obtain final systematic errors 
separately for each decay channel.
The contributions are listed in Table 1.

A significant systematic uncertainty comes from the uncertainty of the 
charged track reconstruction efficiency, which has been evaluated from data.
This uncertainty is estimated to be $2\%$ per track for tracks with
momentum larger than 200 MeV/c.
A special study was performed to determine the momentum dependence
of the reconstruction efficiency 
of the charged low momentum pion ($P^{CM}(\pi_{slow})\,<\,(220-250)$MeV/c)
from the $D^{*+}$ meson decay.
The distribution of the helicity angle, which is strongly correlated
with the slow pion momentum, is compared in MC and data for
$B^0 \rightarrow D^{*+} \pi^-$ decays. Good agreement of the shapes 
of these distributions is obtained.
A similar method is used to estimate the reconstruction 
efficiency of a $\pi^0$ meson produced in $D^{*0}$ decay.
A test of the relative reconstruction efficiencies for low momentum 
$\pi^+$ and $\pi^0$ was performed by comparing two 
inclusive $D^{*+}$ decay modes. The ratio of the branching fractions 
$R_{D^{*+}}= {\it Br}(D^{*+} \rightarrow D^+ \pi^0)
/ {\it Br}(D^{*+} \rightarrow D^0 \pi^+)$ 
was determined in generic $B\bar{B}$ MC and data. Reasonable 
agreement was obtained.
A 10$\%$ systematic uncertainty is added to take
into account the uncertainty in the reconstruction efficiency 
for both charged and neutral slow pions.

The charged kaon and pion identification efficiency is checked 
by fitting the $D^{*+} \rightarrow D^0 \pi^+$,
$D^0 \rightarrow K^- \pi^+$ signal in MC and data.
The fractions of signal events with and without kaon or pion identification 
are compared in the $D^{*+}$ center-of-mass momentum range
from 1.0 to 2.5 GeV/c. 
This systematic error is found to be around $1\%$ per particle.
A similar procedure is applied to check the $K^0_S$ selection 
efficiency, where the $D^{*+} \rightarrow D^0 \pi^+$,
$D^0 \rightarrow K^0_S \pi^+ \pi^-$ decay channel is used.
The systematic uncertainty obtained is $3\%$.

The non-resonant background under the $K^{*0}$ signal
is estimated from a fit using a relativistic Breit-Wigner function
for the signal and a linear function to describe the 
non-resonant background.
The mass and width of the $K^{*0}$ are
in good agreement with the nominal values.
The non-resonant contribution under the $K^{*0}$ is found to be
($2\pm2)\,\%$. No correction is applied and
an uncertainty of $4\%$ was added for this background source.
The efficiency of the $D$ meson mass and vertex constrained kinematic 
fit is close to 100$\%$ with about a $1\%$ uncertainty.
The uncertainties in the $D$ and $D^*$ meson decay branching fractions
were also taken into account \cite{pdg}.

The probability of finding more than one $B$ candidate per event 
is generally not large and is well reproduced by MC.
The fraction of multiple entries varies from 2$\%$ for channels 
with only charged tracks in the final state, to 17$\%$ for channels with 
a $D^{*0}$. The uncertainty in the signal yield
due to this effect is estimated to be $4\%$ for the decay modes with a 
$D^{*0}$ and $2\%$ for other decay modes.           

The background under the $B$ signal is separated into the
combinatorial background, which is removed by the fit procedure,
and peaking backgrounds, where the content of the final particles is 
the same as in the studied channel, but a $D$ meson was not formed.
The systematic uncertainty of the combinatorial background 
subtraction is estimated by varying the function that describes
the background shape. This uncertainty is smaller than 5$\%$.
The width of the signal Gaussian is also varied inside the acceptable 
range and this systematic error is found to be in the range (3--5)$\%$. 
The uncertainty in the estimate of the peaking
background is found to be approximately $5\%$ for the 
$\doko$ and $\dokst$
decay channels and significantly smaller for other channels, if 
this background is evaluated from $D$ meson sideband studies.

The MC distributions of $K^- K^{*0}$ and $K^- K^0_S$ mass were adjusted
to describe the data.
The shapes of these distributions are varied within their experimental 
uncertainties.
For modes with $D^{*0}$ and $D^{*+}$ mesons the polarization is varied
from longitudinal to transverse. 
The uncertainty in the determination of the number of $B\bar{B}$ pairs
for the full data sample is 1$\%$.
Finally, the MC statistical uncertainties
are also included in the systematic uncertainty.
The overall systematic uncertainty ranges from 15 to 19$\%$
for modes with $D$ mesons and from 19 to 22$\%$ for
modes with $D^*$ mesons (Table 1).

\renewcommand{\arraystretch}{1.1}
\begin{table}[h!]
\vspace{0.3cm}
\caption{The systematic uncertainties in the branching fraction
calculations.}
\vspace{0.3cm}
\begin{center}
\begin{tabular}{|ll|c|}
\hline
1. & Reconstruction efficiency &   \\
 & a) Charged track (per track) & 2$\%$ \\
 & b) Slow $\pi^+$ from $D^{*+}$  & 10$\%$ \\
 & c) Slow $\pi^0$ from $D^{*0}$  & 10$\%$ \\
 & d) $K^0_s$, $K^{*0}$, $D^0$ and $D^+$ mesons & (1-3)$\%$ \\
2. & $\pi^{\pm}/K^{\pm}$ identification (per particle)& 1$\%$ \\
3. & Background subtraction &   \\
  & a) Combinatorial background shape & (1-5)$\%$  \\ 
  & b) $B$ signal width &  (3-5)$\%$ \\ 
  & c) Combinatorial background under $D^0$ & (1-5)$\%$ \\
  & d) Non-resonant background under $K^{*0}$ & 4$\%$ \\ 
4. & MC model & \\ 
  & a) $D^{*0}$ and $D^{*+}$ polarization & (4-8)$\%$ \\ 
  & b) $K^-K^{0(*)}$ mass shape & (2-3)$\%$ \\
  & c) $B$ candidate multiple entries & (2-4)$\%$ \\ 
  & d) MC statistical error & (3-6)$\%$  \\
5. & $D$ and $D^*$ branching fractions & (2-7)$\%$ \\ 
6. & Number of $B$ mesons & 1$\%$ \\ \hline
\multicolumn{2}{|c|} { Total: \ \ \ \ \ \ \ \ \ \ \ \  $D$ \ modes} & $\sim (15-19)\%$   \\ 
\multicolumn{2}{|c|} { Total: \ \ \ \ \ \ \ \ \ \ \ \  $D^*$ modes} & $\sim (19-22)\%$ \\ \hline
\end{tabular}
\vspace{0.1cm}
\end{center}
\label{tab:syst}
\end{table}

\section{Results}

After applying all the selection requirements,
the signal yields for the eight 
decay channels are obtained from fits to the $\Delta E$ 
distributions shown in Fig.~2. 
Because the $\Delta E$ and $M_{\rm bc}$ parameters are almost
independent, an additional selection $M_{\rm bc} > 5.272$\,GeV/c$^2$
was applied to suppress background.
The $B$ signal is described by a Gaussian 
function with a width fixed from MC.
The peak position is fixed to zero if the signal statistical significance
is less than 4$\sigma$. For decay modes with large numbers of events,
the peak position is allowed to float.
The background is described by a first-order polynomial, which
fits events in the $B$ meson mass sideband
$(5.2<M_{\rm bc}<5.26)\,$GeV/c$^2$, shown
as the hatched histograms in Fig.~2.
The backgrounds under the signals are well
reproduced by the $B$ mass sideband events.
The fit interval $-0.12 <\Delta E<0.2\,$GeV is chosen
to avoid reflections from modes with an additional
soft pion, in particular from $D^*$ mesons.
The $M_{\rm bc}$ distribution for the range $|\Delta E| < 25\,$MeV
was fitted by a Gaussian function with a fixed
width to describe the signal
and by the so-called ARGUS background function \cite{argus}
with floating exponential
and normalization parameters to describe the background.
The peak position is fixed to the value 5.2795$\,$GeV/c$^2$
if the signal statistical significance is less than 4$\sigma$.
The shape of the background is chosen to be flat if the population
outside the signal region is small.
The signal yields obtained from the $\Delta E$ and $M_{\rm bc}$ fits
are listed in Table 2 and found to be in good agreement.

\begin{figure}[t!]
\vspace{0.1cm}
\begin{center}
\epsfig{file=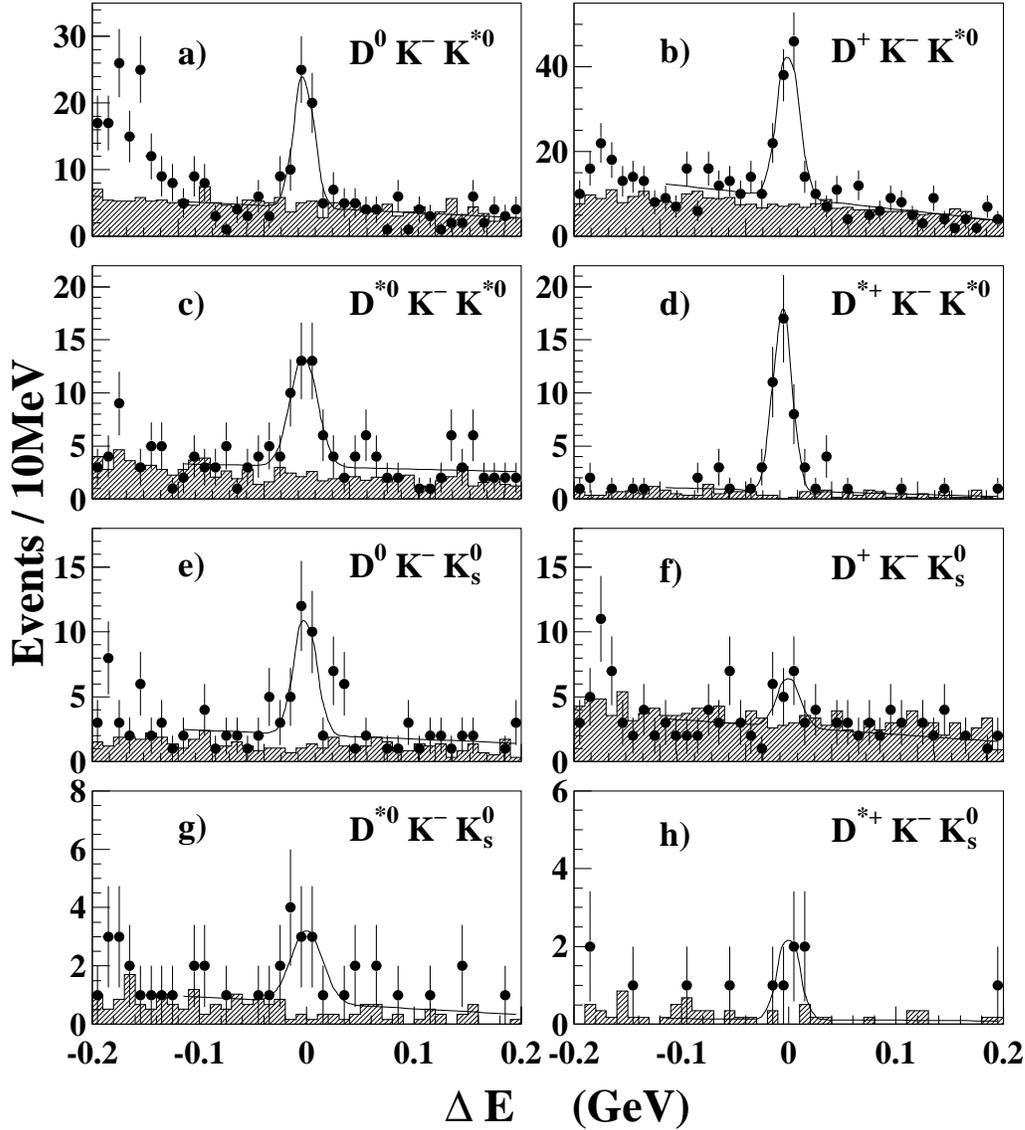,width=15.6cm,height=15.6cm}
\end{center}
\vspace{0.1cm}
\caption{The $\Delta E$ distributions for eight $B$ meson decay channels
(see Table 1). The points with error bars are the data, the curves
show the results of the fits described in the text. The hatched 
histogram shows the $\Delta E$ distribution for events in the $B$ mass 
sideband $(5.2\,<\,M_{\rm bc}\,<\,5.26)\,$GeV/c$^2$.}
\label{deltae}
\end{figure}

\renewcommand{\arraystretch}{1.2}
\begin{table}[h!]
\vspace{0.3cm}
\caption{The signal yields (obtained from the $\Delta E$ and $M_{\rm bc}$ fits),
efficiencies (for $K\pi$  and $K\pi\pi\pi$ decay modes), 
branching fractions, upper limits
and significances (the probabilities of a background fluctuation,
measured in equivalent $\sigma$ units) for the studied decay modes. The 
branching fractions are calculated using the yield from the 
$\Delta E$ fit.}
\vspace{0.3cm}
\begin{center}
\begin{tabular}{|l|c|c|c|c|}
\hline
   & Yield & Eff.($\%$) & Br. fractions, & Signif. \\
 Decay modes & $\Delta E$ / $M_{\rm bc}$ & K$\pi$/K$\pi\pi\pi$ & upper limits (10$^{-4}$) & $\sigma$ \\ \hline \hline
 $\dokst$ & 46.7 $\pm$ 8.2  & 7.71 & 7.5 $\pm$ 1.3 $\pm$ 1.1 & 8.0 \\
          & 46.4 $\pm$ 7.4  &  -   &                         &     \\
\hline
 $\dpkst$ & 87.7 $\pm$ 11.4 & 5.23 & 8.8 $\pm$ 1.1 $\pm$ 1.5 & 10.4 \\ 
          & 88.8 $\pm$ 10.2 &  -   &                         &      \\ 
\hline
 $\dsokst$ & 32.8 $\pm$ 7.2  & 2.72 & 15.3 $\pm$ 3.1 $\pm$ 2.9 & 6.7 \\
           & 37.3 $\pm$ 6.9  & 0.99 &                          &     \\
\hline
 $\dspkst$ & 37.5 $\pm$ 6.4  & 3.28 & 12.9 $\pm$ 2.2 $\pm$ 2.5 & 9.5 \\
           & 38.6 $\pm$ 6.3  & 1.05 &                          &     \\
\hline
 $\dokoo$  & 23.7 $\pm$ 5.9  & 10.25 & 5.5 $\pm$ 1.4 $\pm$ 0.8 & 5.5 \\
           & 28.1 $\pm$ 5.8  &  -   &                         &     \\
\hline
 $\dpkoo$  & 10.3 $\pm$ 5.0  & 6.62 & 1.6 $\pm$ 0.8 $\pm$ 0.3 & 2.6 \\
           &  5.8 $\pm$ 4.5  &  -   &   $<3.1\ (90\%\ CL)$     &     \\
\hline
 $\dsokoo$ &  9.1 $\pm$ 3.9  & 3.36 & 5.2 $\pm$ 2.7 $\pm$ 1.2 & 2.5 \\
           & 10.5 $\pm$ 3.5  & 1.26 &   $<10.6\ (90\%\ CL)$    &     \\
\hline
 $\dspkoo$ & 5.4  $\pm$ 2.5  & 4.46 & 2.0 $\pm$ 1.5 $\pm$ 0.4 & 2.5 \\
           & 5.6  $\pm$ 2.7  & 1.49 & $<4.7\ (90\%\ CL)$       &     \\ 
\hline
\end{tabular}
\end{center}
\vspace{0.3cm}
\label{tab:syst}
\end{table}

Significant signals are observed in the channels with
$K^-K^{*0}$ subsystems and in the $\doko$ decay mode (Table 2).
The signals in the $\dpko$, $\dsoko$ and $\dspko$ decay modes 
are not statistically
significant, and 90\% confidence level (CL) upper limits 
for their branching fractions are also given in Table 2.
We calculate the upper limits assuming a Gaussian distribution 
of the statistical error, and then inflate the limit by one unit 
of the systematic error.
The branching fractions for the decays of the intermediate
$D$, $D^*$, $K^0$ and $K^{*0}$
states were not included in the efficiencies and their values were taken
from Ref.\cite{pdg}.  
An equal production rate for the neutral and charged $B$ mesons
is assumed in the calculation of the branching fractions.

For further analysis the events from the $B$ meson signal region 
are selected by the requirements
$|\Delta E| < 25\,$MeV and $M_{\rm bc} > 5.272$\,GeV/c$^2$. 
The $K^- K^{*0}$ invariant mass distribution after
background subtraction and efficiency correction
is shown in Fig.~3a for the combined signal region events for the
four corresponding $B$ decay modes. 
The efficiencies as a function of $K^- K^{*0}$ mass
are flat within the errors, except for the bin
where the $D_s^-$ signal is removed.
The background under the $B$ signal is modeled using
the $M_{\rm bc}$ sideband.

\begin{figure}[t!]
\vspace{0.1cm}
\begin{center}
\epsfig{file=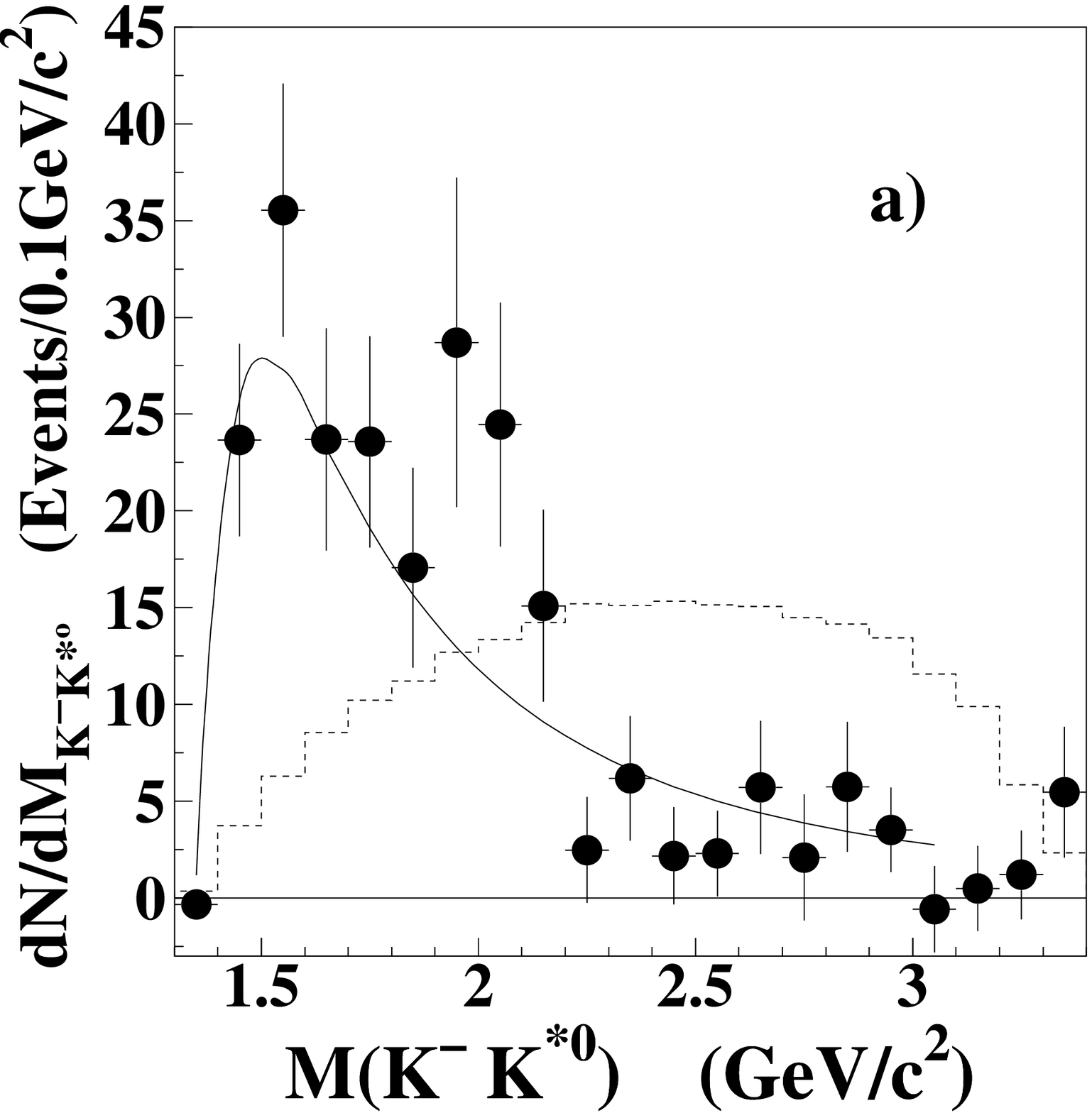,width=7.cm,height=7.cm}\epsfig{file=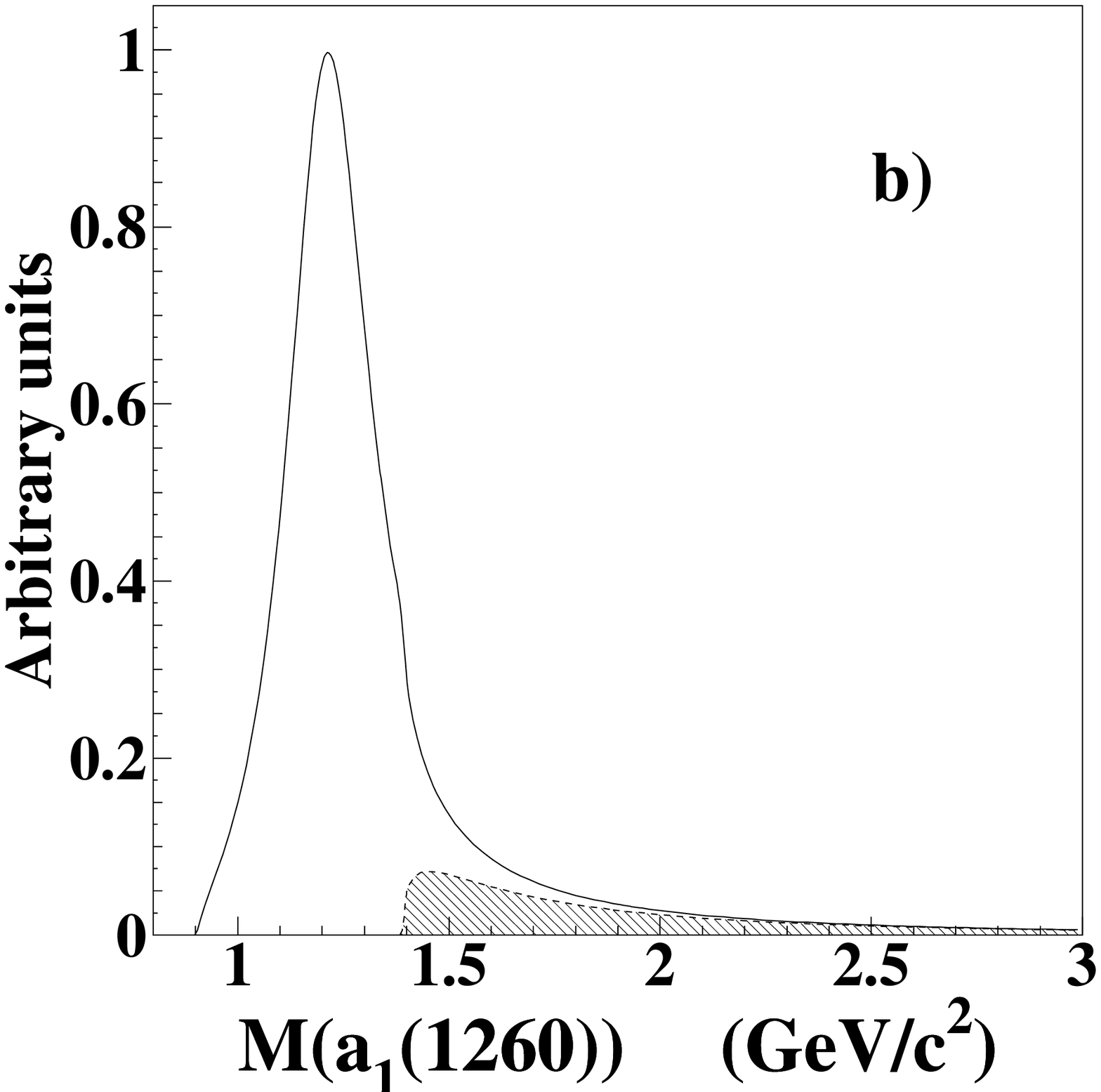,width=7.cm,height=7.cm}
\end{center}
\vspace{0.1cm}
\caption{a) The $K^- K^{*0}$ mass distribution after background
subtraction and efficiency correction for the
four $B$ decay modes combined. The curve shows
the fit to the relativistic Breit-Wigner function
described in the text. The dashed histogram shows the distribution
for the three-body phase space $B$ decay.
b) The shape of the $a_1$ resonance invariant
mass for the $a_1^-(1260) \rightarrow \rho^0 \pi^-$ decay channel 
is shown by the solid line.
The dashed line (hatched area) shows the shape of the $K^- K^{*0}$ mass 
distribution for the $a_1^-(1260) \rightarrow K^- K^{*0}$ decay channel.}
\label{M23a}
\end{figure}

Simulation of the three-body decay $B \to D^{(*)} K^- K^{*0}$ according
to phase space results in the
$K^- K^{*0}$ mass distribution shown by the dashed histogram in Fig.~3a.
It is clear that the
observed shape and an enhancement in the low $K^- K^{*0}$ mass region
rule out any significant phase space contribution.
A MC simulation assuming an intermediate $K^- K^{*0}$ resonance shows that 
the observed low mass enhancement can be explained by the 
quasi-two-body decays $B \rightarrow D^{(*)} a_1^-(1260)$, with
$a_1^-(1260) \rightarrow K^- K^{*0}$.
Figure~3b shows the invariant mass of the
products of the $a_1^-(1260)$ meson decay via its dominant decay mode   
$\rho^0 \pi^-$ (the solid line) and via the decay mode $K^- K^{*0}$
(the hatched area).
The shapes of the distributions were described by relativistic 
Breit-Wigner functions where both decay channels 
contribute to the total width simultaneously \cite{Jackson}.

The $K^- K^{*0}$ mass distribution was parameterized by the sum 
of the Breit-Wigner function for a resonance 
and a phase space function $F_{PS}(M_{KK^*})$ with a fixed shape for the 
non-resonant three-body phase space $B$ decay,

\begin{equation}
\frac{dN}{dM_{KK^*}} = \frac{f_{a_1} M_{a_1} \Gamma_{part} (M_{KK^*})}
{(M_{a_1}^2-M_{KK^*}^2)^2+M_{a_1}^2 \Gamma^2_{tot}(M_{KK^*})}
\ + \ f_{PS} \cdot F_{PS}(M_{KK^*}) \ \ ,
    \label{eq:bw}
\end{equation}
where the mass dependence of the partial $\Gamma_{part} (M_{KK^*})$ and 
total $\Gamma_{tot}(M_{KK^*})$ decay widths is approximated
assuming a two-body phase space \mbox{$a_1 \rightarrow K K^*$} decay:
\begin{equation}
\Gamma_{part} (M_{KK^*}) = \Gamma_{KK^*} \cdot (q_{KK^*}/q_0) \ , \ \ \ \ \
\Gamma_{tot} (M_{KK^*}) = \Gamma_{a_1} \cdot (q_{\rho\pi}/q_0).
    \label{eq:bwk}
\end{equation}

Here $f_{a_1}$ and $f_{PS}$ are normalization parameters, and $M_{a_1}$ and 
$\Gamma_{a_1}$ are the mass and the 
width of the $a_1$ resonance. 
The function $F_{PS}(M_{K K^*})$ describing the three-body phase space   
distribution obtained by simulation was parameterized by a product
of polynomial and threshold functions.
The $q_{\rho\pi}$, $q_{KK^*}$ and $q_0$ parameters are three-vector 
momenta of each daughter particle in the $a_1$ rest frame. 
The constant $q_0$ is calculated assuming a
$a_1 \rightarrow \rho \pi$ decay with the mass of the $a_1$ resonance fixed 
to its nominal value.
The total width $\Gamma_{tot}(M_{KK^*})$ is approximated by the 
dominant partial width of the $a_1 \rightarrow \rho \pi$ decay mode.

If the $a_1$ mass is fixed to its nominal value
$M_{a_1}=1230\,$MeV/c$^2$ \cite{pdg} and the width is fixed 
to any value in the range $300 < \Gamma_{a_1} < 600$~MeV/c$^2$, the relative 
contribution of the $a_1$ extracted from the fit is almost 100$\%$. 
Introducing an additional interference term in the formula (1) leads 
to a small correction, which does not change the fit results.
If the phase space term is omitted, a fit with 
the resonance width floating gives 
$\Gamma_{a_1}=(460\,\pm\,85)\,$MeV/c$^2$ (Fig.~3a), which agrees well
with the nominal range for the $a_1$ width \cite{pdg}.
The quality of the fit is reasonable: $\chi^2/n.d.f.$ = 29.5/18.
Unfortunately, contributions from non-resonant three-body $B$ decays
that do not have mass distribution with a phase space shape and/or
higher resonances cannot be reliably separated due to the low statistics.
The branching fractions for $B \rightarrow D^{(*)} a_1^-(1260)$
have been measured by CLEO in the 
$a_1 \rightarrow \rho \pi$ mode \cite{cleoa}. Comparing the CLEO measurement 
with our results and assuming purely resonant
production of the $K^- K^{*0}$ system, the relative fraction
$R=Br(a_1^-(1260) \rightarrow K^- K^{*0})/ Br(a_1^-(1260) \rightarrow all)$ 
is found to be within the range (8-15)$\%$ depending on the
$B$ decay mode studied. 
This is larger than the value $(3.3\,\pm\,0.5)\%$
that CLEO obtained by 
studying the three pion mass shape in $\tau$ decays \cite{cleob}.

To study the quantum numbers of the intermediate $K^- K^{*0}$ state, 
the angular distributions of the final state particles were 
examined. First, the helicity angle $\theta_{KK}$
was defined as the angle between the momentum of the $K^-K^{*0}$
system in the $B$ meson rest frame and the momentum 
of the $K^{*0}$ in the $K^-K^{*0}$ system rest frame. The helicity 
angle $\theta_{K^*}$, which was
defined as the angle between the momentum  of $K^{*0}$
in the $K^-K^{*0}$ system rest frame and the momentum  of the
daughter $K^+$ meson in the $K^{*0}$ rest frame, was also studied.

Table 3 gives the expected distributions for these two helicity angles
for the possible quantum numbers of the $K^-K^{*0}$ system in 
$B \rightarrow D K^-K^{*0}$
and $B \rightarrow D^* K^-K^{*0}$ decays. 
One can see that for J$^P=0^-$ and $1^+$ 
the angular distributions for the modes with $D$ and $D^*$ are the same.
However, for modes with a $D^*$ and J$^P=1^-$ the situation becomes more 
complicated and the angular distributions depend on the fraction 
of the longitudinal polarization in the $B$ decay.

\begin{figure}[t!]
\vspace{0.1cm}
\begin{center}
\epsfig{file=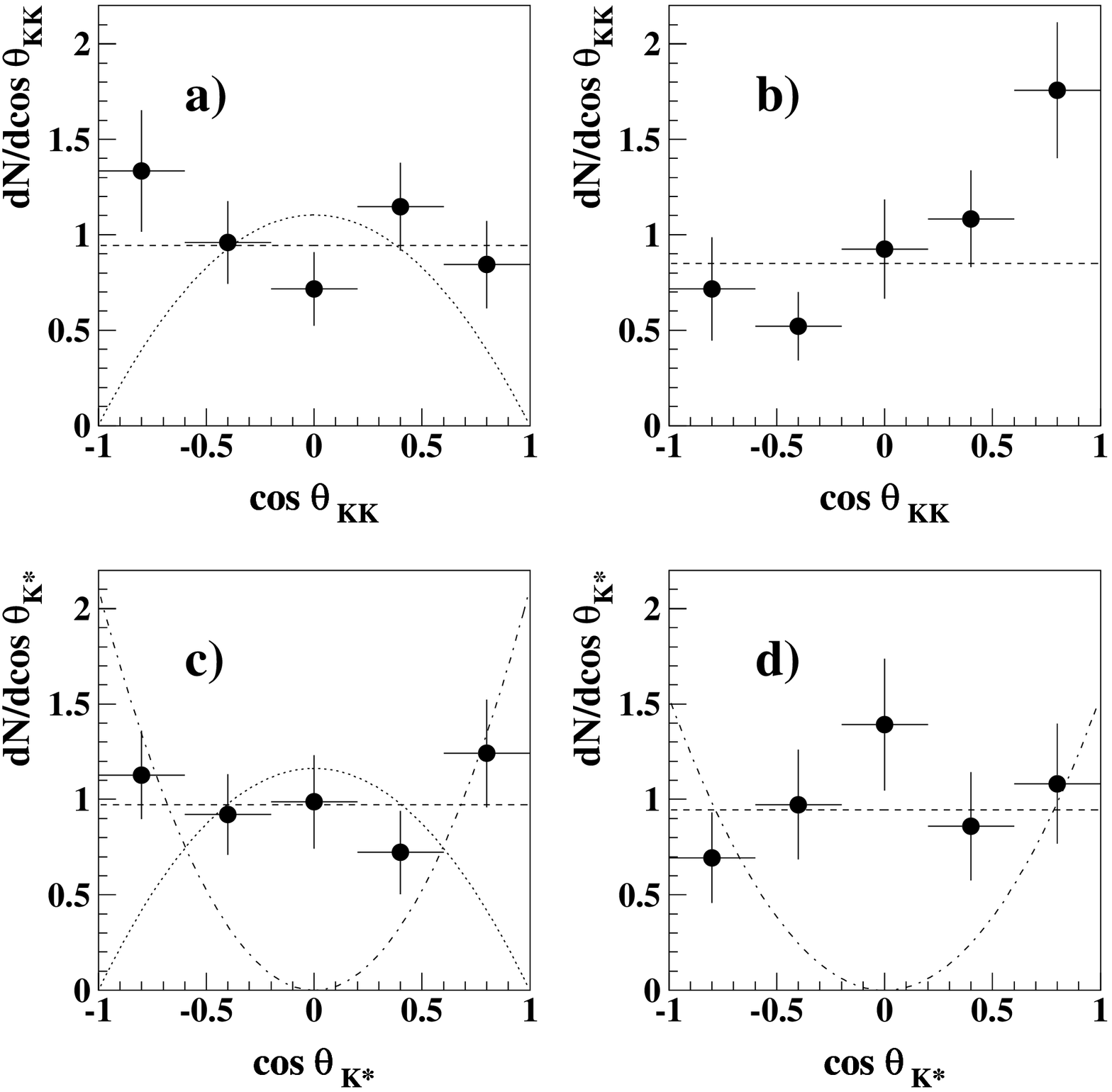,width=13.5cm,height=13.5cm}
\end{center}
\vspace{-0.1cm}
\caption{The cos$\,\theta_{KK}$ distributions: a) for the sum 
of the $\dokst$ and $\dpkst$ modes and b) for the sum of
the $\dsokst$ and $\dspkst$ modes. The cos$\,\theta_{K^*}$ 
distributions: c) for the sum of the $\dokst$ and $\dpkst$ modes 
and d) for the sum of the $\dsokst$ and $\dspkst$ modes.
The curves are fits to a constant (dashed), 
sin$^2 \theta$ (dotted) and cos$^2 \theta$ (dashed-dotted) functions.}
\label{Pol1}
\end{figure}
\vspace{0.1cm}

\renewcommand{\arraystretch}{1.2}
\begin{table}[h!]
\vspace{0.1cm}
\caption{The angular distributions of the $K^-K^{*0}$ system
with J$^P=0^-, 1^-, 1^+$. The values of $\chi^2/n.d.f.$ 
were obtained from fits to the experimental angular 
distributions (see Fig.~4).}
\vspace{0.2cm}
\begin{center}
\begin{tabular}{|l|c|c|c|c|c|}
\hline
     & \multicolumn{2}{|c|}{$D$ meson modes} & \multicolumn{2}{|c|}{$D^*$ meson modes} & Sum \\ \cline{2-5}
 J$^P$ & $\theta_{KK}$ & $\theta_{K^*}$ & $\theta_{KK}$ & $\theta_{K^*}$ & $\chi^2/n.d.f.$ \\ \hline
 0$^-$ & const & cos$^2 \theta_{K^*}$ & const & cos$^2 \theta_{K^*}$ & 71.7 / 16 \\
 1$^-$ & sin$^2 \theta_{KK}$ &  sin$^2 \theta_{K^*}$ & - & - & 37.3\, / \ 8  \\
 1$^+$ & const & const & const & const & 20.7 / 16 \\ \hline
\end{tabular}
\end{center}
\vspace{0.1cm}
\label{tab:syst}
\end{table}

Figure 4 shows the cos$\,\theta_{KK}$ and cos$\,\theta_{K^*}$
distributions for the first four $B$ decay mode combinations 
after background subtraction and efficiency correction.
The angular resolution in these measurements is significantly smaller
than the bin size.
No significant deviation from the flat dependence is observed for all
studied angular distributions, except for the bin
at high cos$\,\theta_{KK}$ in Fig.~4b.
The results of the fits are also shown in Fig.~4.
The $\chi^2/n.d.f.$ values given in the last column of Table 3
were obtained by summing the values for the four individual fits. 
The fit results are consistent with a pure J$^P$=1$^+$ state for 
the $K^-K^{*0}$ system 
and rule out pure J$^P$=0$^-$ or J$^P$=1$^-$ states. 
It has to be stressed that only decay 
modes with $D$ mesons are used to test the angular distributions
for J$^P$=1$^-$ case.
The fit results support the hypothesis that the intermediate 
$a_1^-(1260)$ resonant state decays to the $K^-K^{*0}$ final state.

In the four decay modes to $K^-K^0_S$ final states, a
clear signal is only observed in the $\doko$ decay mode (Fig.~2e),
the enhancements in the other three decay modes (Figs.~2f,g,h) each have
less than 3$\sigma$ significance.
The $K^- K^0_S$ mass spectrum for the $\doko$ decay mode
after background subtraction and efficiency correction is shown 
in Fig.~5a, where the $D^0\rightarrow K^-\pi^+$ decay channel is used.
The mass distribution peaks near the $K^- K^0_S$ mass threshold;
the fraction of $\doko$ signal events in the 
mass range $M(K^- K^0_S) < 1.3\,$GeV/c$^2$ is $55\%$.
The distribution for the three-body phase space $B$ decay is
shown by the dashed histogram.

The efficiency corrected and background subtracted
helicity angle distribution for the $K^- K^0_S$ system 
is shown in Fig.~5b. The angle $\theta_{KK}$ is defined 
similarly to that for the $K^- K^{*0}$ system, where the $K^0_S$
is substituted for the $K^{*0}$.
This distribution is fitted to the form,
\begin{equation}
\frac{dN}{d{\rm cos}\,\theta_{KK}} = A \left( R_L \,
{\rm cos}^2\theta_{KK} \, + \, 
(1-R_L) \, {\rm sin}^2\theta_{KK} \right) \ \ ,
    \label{eq:costheta}
\end{equation}
where $R_L = \Gamma_L/\Gamma$ is the fraction of longitudinal 
polarization and $A$ is a free normalization parameter. 
A fit to the functional form (3) gives
the value $R_L = 0.97 \pm 0.08$, which is typical of a P-wave
decay of a J$^P=$1$^-$ system to two pseudoscalar mesons.

\begin{figure}[t!]
\vspace{0.1cm}
\begin{center}
\epsfig{file=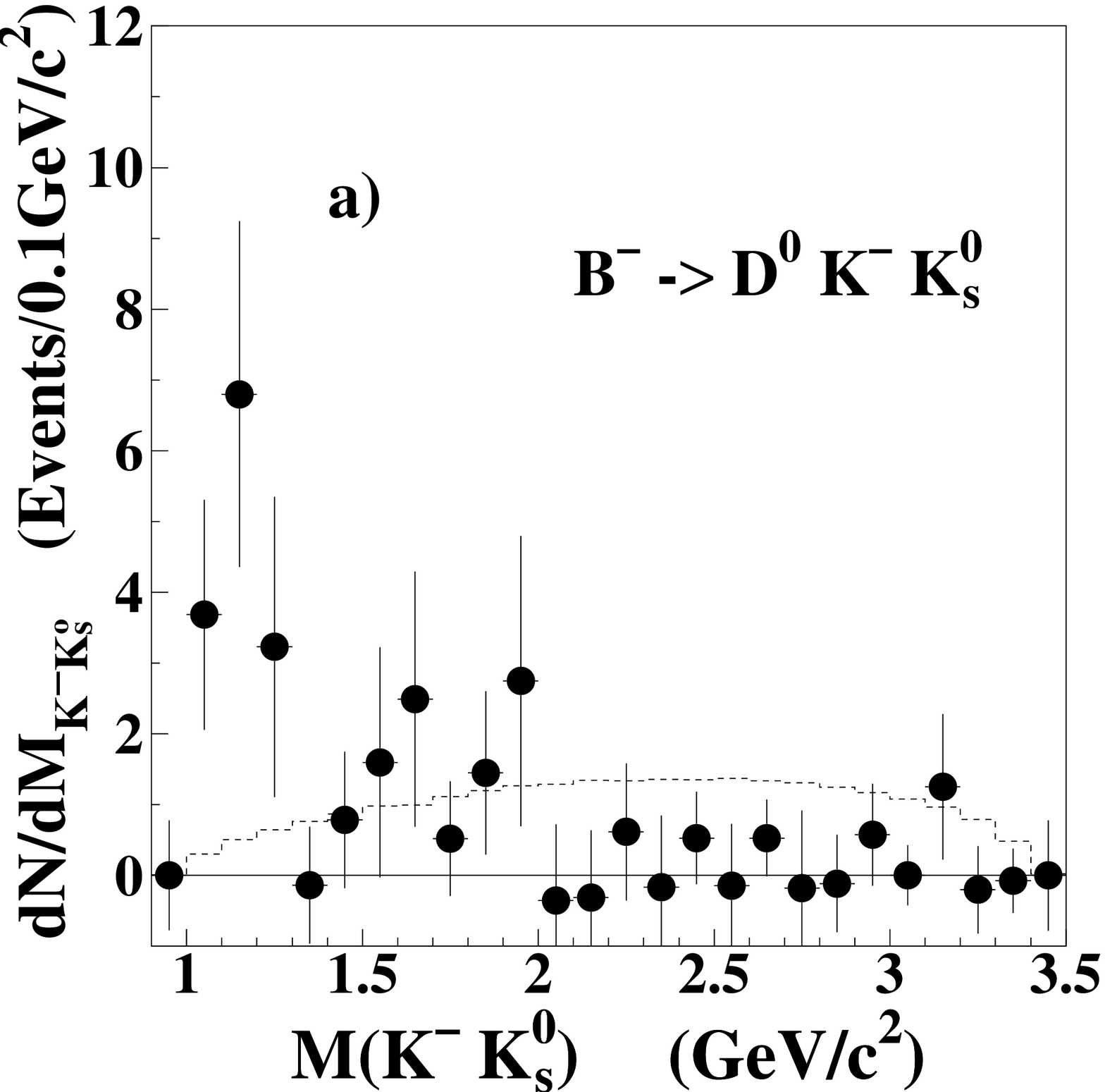,width=7.cm,height=7.cm}\epsfig{file=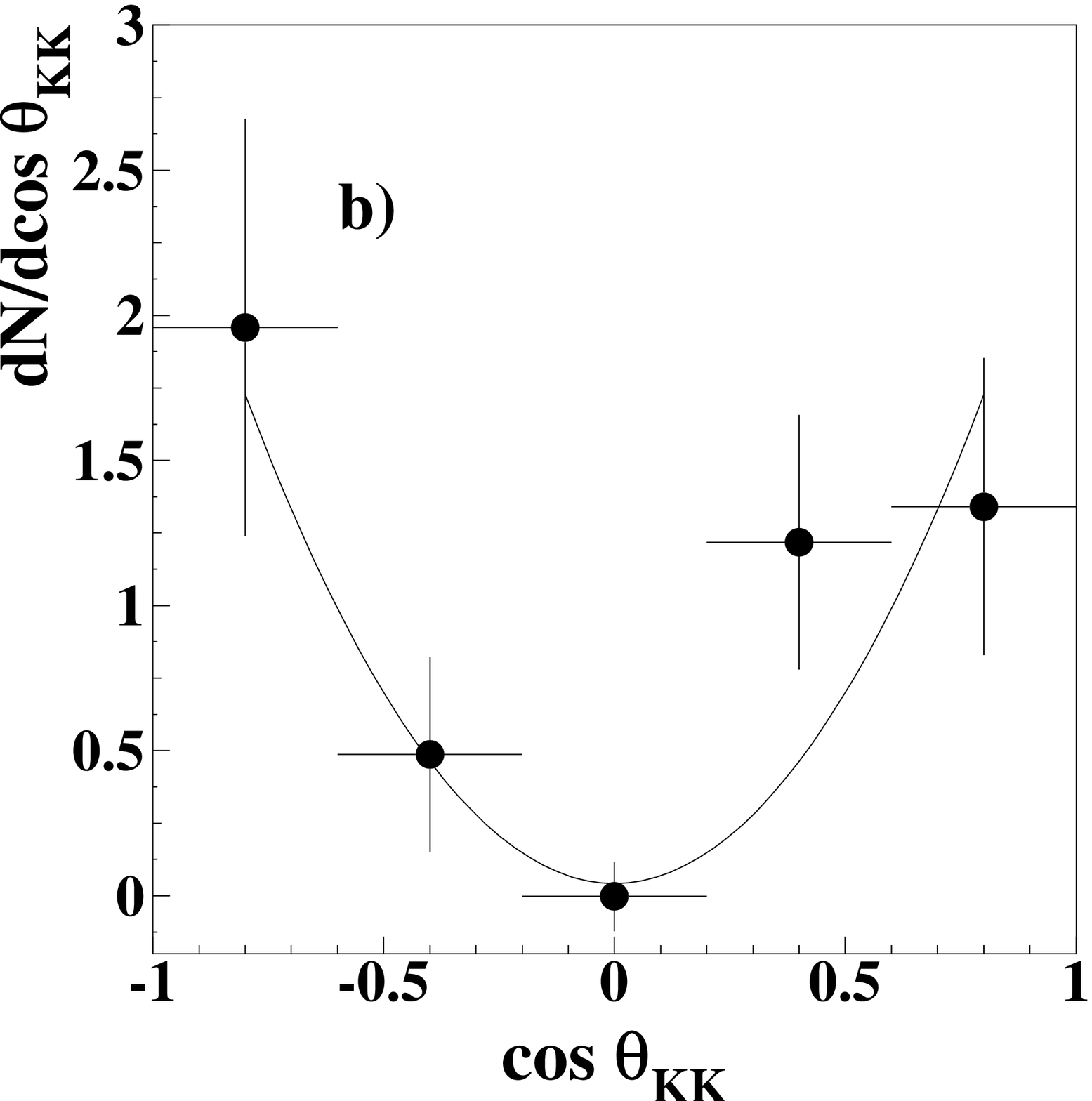,width=7.cm,height=7.cm}
\end{center}
\vspace{0.1cm}
\caption{The fully corrected a) $K^- K^0_S$ invariant mass and b)
$K^- K^0_S$ helicity angle distributions.
The data are from the $\doko$ decay mode with 
$D^0 \rightarrow K^-\pi^+$. The dashed histogram shows
the expected distribution for three-body phase-space $B$ decays.
The curve shows the result of the fit to the function described in the text.}
\vspace{0.2cm}
\label{M23a}
\end{figure}

The observed mass distribution is expected to include
interfering contributions from
the $\rho(770), \rho(1450)$ and $\rho(1700)$ resonances and follow 
that for the $K \bar{K}$ system produced in the vector isovector state 
in other processes like $e^+e^-$ annihilation into two kaons
or $\tau^-$ lepton decays to $K^- K^0$. Although rather precise
measurements of the cross sections exist for the processes
$e^+e^- \to K^+K^-$ \cite{kch} and $e^+e^- \to K^0_LK^0_S$ \cite{kne}, 
the final states observed in $e^+e^-$ annihilation are a mixture of 
isovector and isoscalar contributions and their separation requires 
a complicated model-dependent analysis. $\tau$ lepton decays are free 
of this disadvantage since in this case the produced $K^- K^0$ system 
always possesses the necessary quantum numbers. CLEO \cite{cleoc}
studied this mass distribution in $\tau$ decays and concluded
that the observed $K^- K^0_S$ mass distribution could qualitatively 
be described by a $\rho$-like intermediate mechanism producing
an enhancement close to the $K^- K^0_S$ threshold. 
  
The measured branching fractions for the decay modes with 
the $K^- K^0_S$ system are of the order of $(2-4)\,\%$ 
relative to those of the respective $B \rightarrow D^{(*)}\rho^-(770)$ 
decay modes with $\rho^-(770) \to \pi^-\pi^0$.
These values are at least 5 times larger than one would na\"{\i}vely estimate,
using $\tau$ decay branching fractions corrected by phase space 
factors \cite{taurho},
even taking into account positive interference of the $\rho(770)$ and
$\rho(1450)$ resonances.

\section{Conclusions}

In summary, eight $\allc$ decay modes have been examined for the
first time and significant signals have been found for five of them.
The branching fractions and upper limits are obtained.
Significant signals 
are observed for the four modes with a $K^- K^{*0}$ system.
The $K^- K^{*0}$ mass spectrum is peaked near threshold
and smoothly decreases with increasing $K^- K^{*0}$ mass. 
The angular dependences indicate that the $K^- K^{*0}$ system has 
J$^P$=1$^+$.
The observed behavior can be interpreted as production of an intermediate 
$a_1^-(1260)$ resonance that then decays to $K^- K^{*0}$. 

A clear signal is also found in the $\doko$ decay mode. The $K^- K^0_S$
mass spectrum for this signal is also peaked
near the $K^- K^0_S$ mass threshold.
The angular dependence shows a cos$^2 \theta_{KK}$ behavior
indicating that the $K^- K^0_S$ system has J$^P$=1$^-$.
Some enhancements in the signal regions are also seen in the 
$\dsoko$, $\dpko$ and $\dspko$ decay modes, 
but with limited statistical significance.
A non-resonant three-body contribution,
which is not described by phase space, cannot easily be ruled out for
any of these decay modes.

\section{Acknowledgment}

We wish to thank the KEKB accelerator group for the excellent
operation of the KEKB accelerator.
We acknowledge support from the Ministry of Education,
Culture, Sports, Science, and Technology of Japan
and the Japan Society for the Promotion of Science;
the Australian Research Council
and the Australian Department of Industry, Science and Resources;
the National Science Foundation of China under contract No.~10175071;
the Department of Science and Technology of India;
the BK21 program of the Ministry of Education of Korea
and the CHEP SRC program of the Korea Science and Engineering Foundation;
the Polish State Committee for Scientific Research
under contract No.~2P03B 17017;
the Ministry of Science and Technology of the Russian Federation;
the Ministry of Education, Science and Sport of the Republic of Slovenia;
the National Science Council and the Ministry of Education of Taiwan;
and the U.S.\ Department of Energy.

\newpage

\end{document}